\documentclass{aa}  

\usepackage{graphicx}
\usepackage{txfonts}
\usepackage{hyperref}
\usepackage{lscape}
\usepackage{longtable}
\hypersetup{breaklinks=true,colorlinks=true,linkcolor=blue,citecolor=blue,urlcolor=blue}

\begin{document}

   \title{Galaxy mass-size segregation in the cosmic web from the CAVITY parent sample}
   \titlerunning{ }


  \author{I. P\'erez\inst{1,2}
\and 
L. Gil\inst{1}
\and
A. Ferr\'e-Mateu\inst{3,4}
\and
G. Torres-R\'ios\inst{1}
\and
A. Zurita\inst{1,2}
\and
M. Argudo-Fern\'andez\inst{1,2}
\and
B. Bidaran\inst{1}
\and
L. S\'anchez-Menguiano\inst{1,2}
\and
T. Ruiz-Lara\inst{1,2}
\and
J. Dom\'inguez-G\'omez\inst{1}
\and
S. Duarte Puertas\inst{1,2,5}
\and
D. Espada\inst{1,2}
\and
J. Falc\'on-Barroso\inst{3,4}
\and
E. Florido\inst{1,2}
\and
R. Garc\'ia-Benito\inst{6}
\and
A. Jim\'enez\inst{1}
\and
R. F. Peletier\inst{7}
\and
J. Rom\'an\inst{8}
\and
P. S\'anchez Alarc\'on\inst{3}
\and
P. S\'anchez-Bl\'azquez\inst{8}
\and
P. V\'asquez-Bustos\inst{1}
 }
\institute{Departamento de F\'isica Te\'orica y del Cosmos, Universidad de Granada, Campus Fuentenueva, Edificio Mecenas, E-18071, Granada, Spain. \email{isa@ugr.es}
\and
Instituto Carlos I de F\'isica Te\'orica y Computacional, Facultad de Ciencias, E-18071 Granada, Spain
\and
Instituto de Astrof\'isica de Canarias, V\'ia L\'actea s/n, 38205 La Laguna, Tenerife, Spain
\and
Departamento de Astrof\'isica, Universidad de La Laguna, 38200 La Laguna, Tenerife, Spain
\and
D\'epartement de Physique, de G\'enie Physique et d’Optique, Universit\'e Laval, and Centre de Recherche en Astrophysique du Qu\'ebec (CRAQ), Québec, QC, G1V 0A6, Canada
\and
Instituto de Astrof\'isica de Andaluc\'ia - CSIC, Glorieta de la Astronom\'ia s.n., 18008 Granada, Spain
\and
Kapteyn Astronomical Institute, University of Groningen, Landleven 12, 9747 AD Groningen, The Netherlands
\and
Departamento de F\'isica de la Tierra y Astrof\'isica \& IPARCOS, Universidad Complutense de Madrid, E-28040, Madrid, Spain
 }

   \date{Received Month Day, Year; accepted Month Day, Year}


  
  \abstract
  {The mass-size relation is a fundamental galaxy scaling relation intimately linked to the galaxy formation and evolution. The physical processes involved in galaxy growth will leave their particular imprint in the relation between the stellar or total mass and the galaxy size.}
  {We aim at exploring the effect of the large-scale environment on the stellar mass-size relation using the samples and some added-value products of the Calar Alto Void Integral-field Treasury surveY (CAVITY) collaboration.}
  {With this goal in mind, we analyse the Petrosian R50 and R90 radii from SDSS DR16 images of a sample of $\approx$ 14000 galaxies inhabiting cosmic voids, filaments/walls, and clusters, with stellar mass range between $\rm 10^{8.5} - 10^{11} M_\odot$. We investigate the mass-size relation with respect to the galaxy morphology, and  with the star formation history (SFH), parametrised by different time-scales ($\rm T_{50}$, $\rm T_{70}$, and $\rm T_{90}$). }
   {We find that, on average, early-type galaxies in voids are approximately 10-20\% smaller than their counterparts in denser environments, such as filaments, walls, and clusters, regardless of when they assembled their mass. Additionally, evidence suggests that the mass-size relation for the more massive void galaxies within the early-type sample has a shallower slope compared to galaxies in denser large-scale environments. In contrast, early-type galaxies in filaments, walls, and clusters exhibit a more consistent mass-size distribution. For stellar masses $\rm log(M_\star / \mathrm{M_\odot}) = 9 - 10.5$, late-type cluster galaxies are smaller and more concentrated than those in lower-density environments, such as filaments, walls, and voids, while void and filament/wall galaxies exhibit similar size and concentration values. However, for galaxies with masses above $\rm 10^{10.5} M_\odot$, the sizes of void galaxies become comparable to those in clusters. The trend of smaller low-mass cluster galaxies is primarily driven by galaxies with $\rm T_{50}$ greater than 7~Gyr.}
   {We conclude that the large-scale environment influences the mass-size relation of galaxies. Assuming that early-type galaxies undergo two growth phases, we find that they primarily grow their mass during the first phase of formation. In voids, the subsequent size growth from minor mergers is less pronounced, likely due to slower evolution and reduced minor merger activity, or because the void environment inherently has fewer accretion events, or a combination of both. The change in the slope for high-mass void galaxies suggests a lower rate of minor accretion. This trend is also evident in late-type void galaxies with masses above $\rm \approx 10^{10.5} M_\odot$, where minor mergers contribute to their size growth. In contrast, late-type quenched cluster galaxies have smaller sizes due to interactions within the cluster environment, with early infallers being more strongly affected by these environmental interactions.}

   \maketitle
%

\section{Introduction}

Galaxy scale relations encode the general processes that dominate galaxy formation and evolution.  One of these fundamental relations is the mass-size distribution. The size of galaxies has been found to vary significantly with stellar mass, morphological type, and redshift \citep{2003MNRAS.343..978S,2015MNRAS.447.2603L}. More massive galaxies tend to have larger radius, early-type morphologies tend to be smaller than similar mass late-type galaxies, and galaxies at higher redshift seem to be smaller, and more compact, than their local counterparts. This growth with age seems to be more efficient for more massive early-type galaxies \citep{2003MNRAS.343..978S,2004ApJ...600L.107F,2006MNRAS.373L..36T,2013ApJ...777..155O,2014ApJ...788...28V}. 

The physical processes affecting the size growth must explain these relations. Late-type galaxies are believed to form in an inside-out fashion, where star formation begins at the centre and gradually expands outward. As a result, the currently younger, actively star-forming regions are typically found in the outer parts of the galactic discs. This growth might be related to external accretion onto the discs \citep[see for instance,][]{2009ApJ...694..396B}. There is some observational evidence that spiral galaxies of intermediate masses are smaller in clusters than in the field \citep{2010MNRAS.402..282M,2013MNRAS.434..325F}, pointing to a possible effect of the environment on the galaxies size, at least for these intermediate late-type galaxies. Early-type galaxies are thought to grow through early mergers \citep[e.g., ][]{2009ApJ...699L.178N}, however, the debate for early-type galaxies remains as whether these galaxies grow while they live in the red sequence or whether they grow from star-forming objects that then quench, known as the "progenitor bias", \citep{2010MNRAS.401.1099H,2011MNRAS.412L...6B,2011MNRAS.415.3903T,2013ApJ...768L..28H}.Therefore, the size of a galaxy is primarily linked to the mass assembly processes that the galaxy undergoes. 

In a previous study by \citet{2023Natur.619..269D} on the statistical analysis of stellar populations from the the central spectrum of galaxies located in voids, filaments/walls, and clusters, it was demonstrated that galaxies exhibit a bimodality in their star formation histories (SFHs). These SFHs were also characterised by parameters such as $\rm T_{50}$, $\rm T_{70}$, and $\rm T_{90}$, representing the timescales, in Gyr, at which galaxies assemble 50\%, 70\%, and 90\% of their stellar mass, respectively. Some galaxies formed about 30\% of their mass early on ($\approx$ 12 Gyr ago, referred to as short-timescale SFH, or ST-SFH), while others assembled their mass more gradually (referred to as long-timescale SFH, or LT-SFH). This bimodal behaviour was observed regardless of the galaxies' current location in the large-scale structure. The study also concluded that ST-SFH galaxies were less influenced by the large-scale environment during their early stages, but differences became more apparent later in their evolution. In contrast, LT-SFH galaxies showed a clear dependence on their large-scale environment throughout their evolution. For these galaxies, differences in $\rm T_{50}$ between various environments could account for up to 3 Gyr, with void galaxies appearing to assemble their stellar mass more slowly compared to those in denser regions. These differences in the mass assembly time-scales might suggest a difference in the mass-size growth rate for galaxies located in different parts of the large-scale structure. 

Numerical simulations do predict a size segregation with environment for early-type galaxies \citep{2013MNRAS.428..109S} with median sizes increasing by a factor 1.5-3 for the high mass halos (high density environment), mainly due to a larger merger fraction.
Previous studies at low redshift (i.e. z~<~0.15) comparing the mass-size distribution with environment reach very different conclusions. The environment in all the cases is either defined as some Nth number of companions within a certain aperture \citep{2010ApJS..186..427N,2010ApJ...715..606N,2014MNRAS.444..682C,2017ApJ...834...73Y,2023ApJ...957...59Y}, by division between field (note that field is not clearly related here to the large scale structure), and cluster objects \citep{2010MNRAS.402..282M,2013MNRAS.433L..59P,2013ApJ...778L...2C}, or defined by halo mass \citep{2013ApJ...779...29H}. These selections include very few void galaxies, as their contribution in mass in the total mass budget of the Universe is of about 10\% \citep[e.g, ][]{2018MNRAS.473.1195L}, so only surveys targeting void galaxies will be able to fully address these galaxies inhabiting these low density regions of the Universe. Moreover, the mass ranges examined in various studies differ significantly. Some focus primarily on massive galaxies with $\rm log(M_\star / \mathrm{M_\odot}) > 10.5$ \citep{2013ApJ...762...77P, 2013ApJ...779...29H}, while others explore galaxies with masses starting from $\rm log(M_\star / \mathrm{M_\odot}) > 9.5$ \citep{2013ApJ...778L...2C,2014MNRAS.444..682C}. These discrepancies in environmental definitions, mass ranges, and methodology, could explain the differing outcomes observed among research groups. The only work to date specifically looking at galaxies in voids reached the conclusion that late-type galaxies in voids are smaller than the average SDSS disk sizes; however, their conclusion is drawn from the analysis of a sample of 15 objects \citep{2011AJ....141....4K}.  No systematic study on the sizes of galaxies in cosmic voids has been carried out until now.   
 
The Calar Alto Void Integral-field Treasury surveY (CAVITY) project aims to unveil the properties of galaxies inhabiting voids through a wealth of multi-wavelength data, including optical integral field spectroscopy (IFS) \citep[see,][]{2024A&A...689A.213P}. The parent sample from the CAVITY project, along with the control samples used to characterise the final dataset \citep{2023A&A...680A.111D,2023Natur.619..269D}, provides an ideal foundation for investigating the mass-size distribution of galaxies across different large-scale environments, a unique opportunity to focus on the mass-size relation of galaxies in cosmic voids, a relationship that has remained largely unexplored until now.  We will make use of the samples and results published within the collaboration. Since the mass density in filaments and walls is comparable \citep[see, for example,][]{2014MNRAS.441.2923C}, we will henceforth refer to the filament/wall environment simply as "filaments" for practical purposes.

This paper is structured in the following way. Sect.~\ref{sec:sample} presents the sample, Sect.~\ref{sec:methodology} explains the parameters used to characterise the size and the light concentration parameter of the galaxies. In Sect.~\ref{results} we present the analysis and results of the mass-size relation derived from the sample. Sect.~\ref{discussion} and Sect.~\ref{summary} present the discussion, summary and conclusions of the work. Hereafter, our assumed cosmology is a flat $\Lambda$CDM cosmology with $\rm \Omega_{m}=0.3,~\Omega_{\Lambda}=0.7,~and~H_{0}=70~km~s^{-1}~\text{Mpc}^{-1}$. Stellar population parameters are based on a Kroupa initial mass function \citep{2001MNRAS.322..231K}. Magnitudes throughout this text are in the SDSS {\it asinh magnitude} system, which closely approximates but does not precisely match the AB magnitude system \citep{2002AJ....123..485S}.

\section{Sample}
\label{sec:sample}
The sample we analyse in this work was originally presented in \citep{2023Natur.619..269D}. For the sake of clarity, we give below a summary of the sample selection.
For the sample of galaxies in cosmic voids, we started from the 4866 galaxies of the parent sample of the CAVITY project \citep[][]{2024A&A...689A.213P}. This is a well defined catalogue of galaxies inhabiting 15 nearby voids \citep[redshift ranging from 0.005 to 0.05, from][]{2012MNRAS.421..926P}, with void effective radii ranging from 15 to 30 Mpc $\rm h^{-1}$. We then excluded those objects in common with the cluster catalogue from \citet{2017A&A...602A.100T}. Afterwards, we selected the galaxies that lie, in projected distance, within 0.8 void effective radii to ensure their belonging to the void, avoiding the edges. This left a sample of 2744 galaxies in cosmic voids. 

For the filament galaxies, we began with the 109945 objects classified as galaxies in the SDSS, all of which have available central spectra and fall within the same redshift range as the CAVITY parent sample (0.005-0.05). We cross-match this list with the void galaxy catalogue from which the parent sample was drawn \citep{2012MNRAS.421..926P}, as well as with the cluster catalogue from \citet{2017A&A...602A.100T}, defining cluster galaxies as those with 30 or more companions \citep{1989ApJS...70....1A}. This process enabled us to exclude both void and cluster galaxies within the specified redshift range. To optimise computational efficiency, we randomly select a sub-sample of 15000 galaxies from this pool, ensuring that the $\rm g-r$ colour, stellar mass, and redshift distributions are preserved (with a KS-test p-value $>0.95$). We further refined the sample by including only galaxies with reliable morphological classifications.
We took the morphological catalogue from \citet{2018MNRAS.476.3661D} to assign a morphology type to galaxies in the samples. This catalogue contains morphological type (T-Type) and Galaxy Zoo 2 survey \citep[GZ2, ][]{2013MNRAS.435.2835W} structural component classifications (disk/features, edge-on galaxies, bar signature, bulge prominence, roundness and mergers) for $\sim 670000$ galaxies in the SDSS. The classifications are obtained with Deep Learning algorithms trained with the GZ2 and \citet{2010ApJS..186..427N} catalogues. In this work, the authors define $P_\mathrm{bulge}$ as the probability of a galaxy to host a prominent bulge and $P_\mathrm{bar}$ as the probability of a galaxy to contain a strong bar.
We removed from our sample those galaxies that lack a morphological classification, and also those galaxies with non-numerical and negative values for the angular Petrosian radii. Applying this criterion we are left with 1526 void galaxies, 8658 objects in filaments, and 3469 cluster galaxies. 

In this study, we also investigate the mass-size distribution in relation to the way galaxies assemble their stellar mass. The stellar population analysis requires high-quality spectra (S/N $\geq$ 20 as measured in the stellar continuum) to reliably derive stellar population parameters \citep[for a detailed explanation of the quality control applied to the SDSS spectra, see][]{2023Natur.619..269D}. To meet this criterion, we refined our sample to address the SFHs by applying this additional quality constraint, resulting in a set of 620 void galaxies, 3754 filament galaxies, and 1928 cluster galaxies.

\begin{figure*}
\centering 
\includegraphics[width = \textwidth]{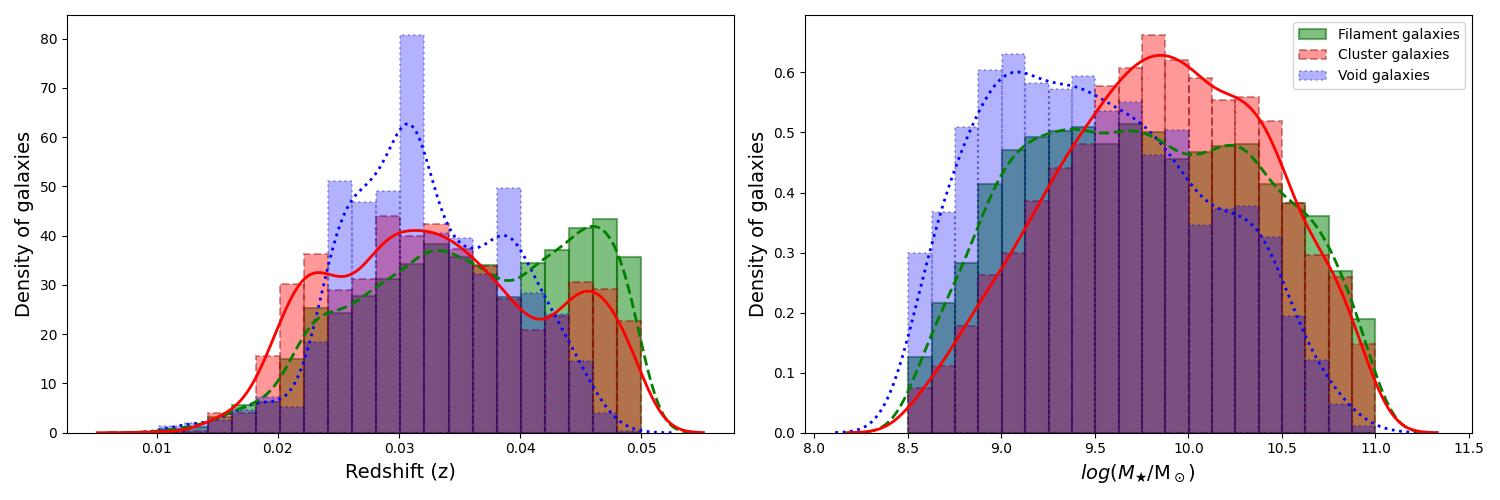} 
\caption{Distribution of stellar mass ($\log(M_{\star}/ \mathrm{M_\odot}$), right panel, and redshift, left panel, for galaxies in voids, filaments, and clusters. Histograms represent the normalised density of galaxies in each large-scale environment, with distinct colors and transparencies for void galaxies (blue), filament galaxies (green), and cluster galaxies (red). KDE (Kernel Density Estimation) plots overlay the histograms to show smoothed distributions. The comparison of the stellar mass distribution highlights that void galaxies tend to populate lower mass ranges, while filament and cluster galaxies extend to higher masses. }
\label{fig:masses} 
\end{figure*}

\begin{figure*}
\centering 
\includegraphics[width =\textwidth]{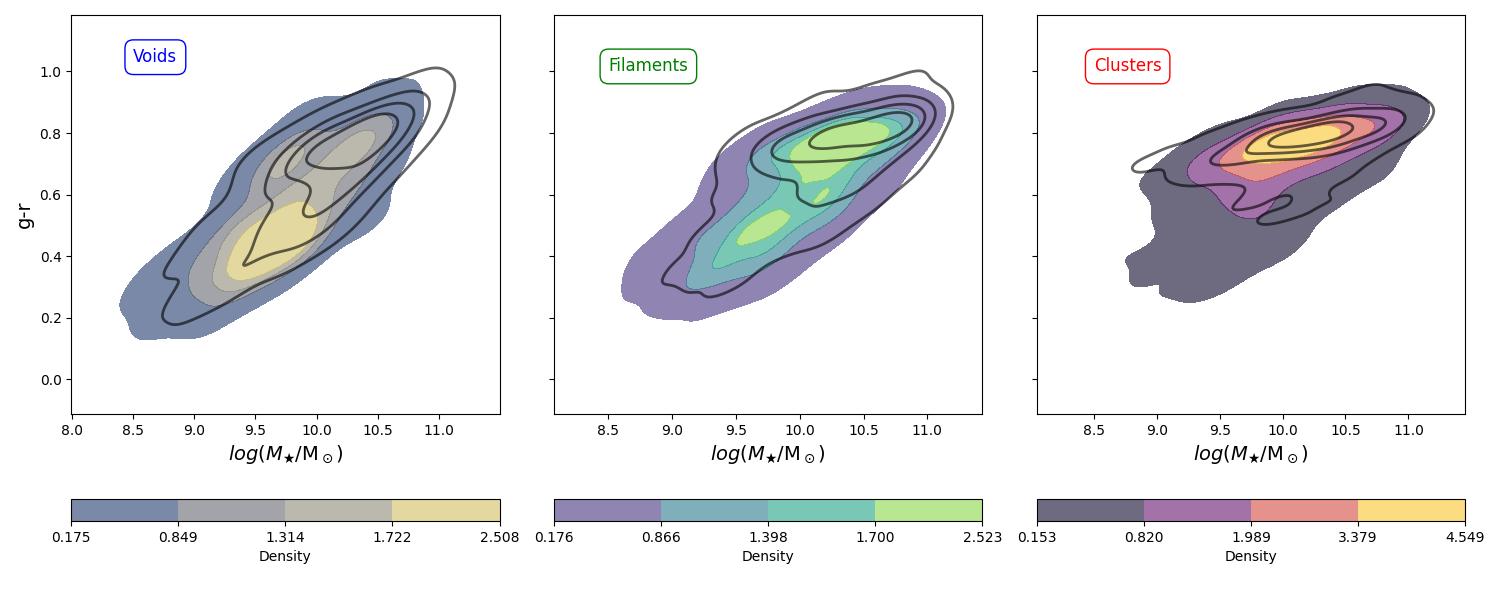}   
\caption{Colour-magnitude diagram for the sample galaxies in the different large-scale environments. The plots show the normalised density of galaxies. The colorbar provides a normalised scale for these density values across the subplots. A distinction between LT-SFH and ST-SFH galaxies has been made, with the colored density maps representing the distributions for LT-SFH systems and the solid contours the density maps for ST-SFH galaxies. Left panel for voids, central panel for filaments and right panel for cluster galaxies. }
\label{fig:sfhdist} 
\end{figure*}

\begin{figure}
\centering 
\includegraphics[width = 0.49\textwidth]{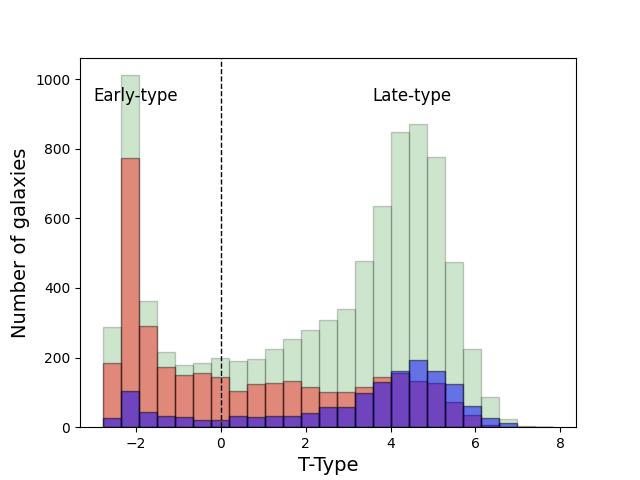} 
\caption{Histogram of the distribution of the morphological types of galaxies in voids (blue), filaments (green), and clusters (red). Morphological types taken from \citet{2018MNRAS.476.3661D}. }
\label{fig:morphology} 
\end{figure}

The right panel of Fig.~\ref{fig:masses} shows the distribution of stellar masses of the final sample. It can be seen from the figure a bimodal distribution of galaxies in filaments (green), the predominance of more massive galaxies in clusters (red), and that void galaxies (blue) tend to populate lower mass ranges. The cut in mass at $\rm log(M_\star / \mathrm{M_\odot}$) > $10^{11}$ is given by the lack of void galaxies at those masses, see Fig.~\ref{fig:sfhdist} for the colour-magnitude distribution of the sample galaxies. Fig.~\ref{fig:morphology} shows the distribution of morphological types for the galaxies in the different large-scale environments, same colour code as in Fig.~\ref{fig:masses}, differentiating between early- and late-type galaxies (lenticular galaxies are included in the early-type group). The left panel of the same figure displays the redshift distribution of galaxies across the different environments. Since the redshift distribution is not uniform in the three environments, we examined the minimum physical size detected within the redshift range and the physical size distribution with redshift for each environment. The minimum physical size detected is consistent across all redshifts, approximately 0.6 kpc, and no significant average size variation with redshift was observed. We also explored limiting the sample at various redshifts, and all conclusions from this study remained valid, though with a reduced sample size. Therefore, we opted to retain the sample covering the full redshift range.
 
\section{Methodology}
\label{sec:methodology}
 
Both theoretical and observational evidence suggests that galaxy size indicators, which link the radial location of the gas surface mass density threshold for star formation to the galaxy's physical size, such as the $\rm R_{1}$ presented by \citet{2020MNRAS.493...87T}, can reduce the scatter in scaling relations involving galaxy size. $\rm R_{1}$ represents the radial position at which the stellar mass density of a galaxy reaches approximately 1$\rm M_{\odot}~pc^{-2}$. This location is chosen to approximate the gas density threshold necessary for star formation. While such indicators would be ideal, they require imaging depth beyond what is available in the SDSS data used in this study.

For practical reasons, and to facilitate comparison with previous work, we use the Petrosian radii, R50 and R90, in the $\rm r$-band as proxies for galaxy size in this study. These radii represent the distances from the center of a galaxy that enclose 50\% and 90\% of the total light in the $\rm r$-band, respectively. While many studies have used the effective radius, there is a subtle but important distinction between the two. The effective radius is generally derived from an analytical fit to the light distribution, making it more sensitive to the shallow surface brightness limits of SDSS images. In contrast, the Petrosian radius is less affected by such limitations and offers a more robust measure for determining galaxy extent \citep[see, e.g.,][]{2024A&A...682A.110B}. However, Petrosian radii are highly dependent on the light intensity profiles of galaxies. On average, early-type galaxies are more concentrated than late-type galaxies. Late-type galaxies, with their bright bulges, bars, and irregular shapes, tend to have less luminosity concentrated near their centers. As a result, the difference between the radii containing 50\% and 90\% of the light is relatively small. In contrast, early-type galaxies exhibit more uniform light distributions, leading to a larger separation between R50 and R90. For this reason, we distinguish between early- and late-type galaxies in our analysis and use the R90/R50 ratio as a proxy for light concentration throughout this work. Summarising, we use the Petrosian R50 and R90 from the SDSS DR16 images \citep{2020ApJS..249....3A} of the sample described in Sect.\ref{sec:sample} to characterise the size of galaxies.

Galaxies with T-Type $\leq 0$ are classified as early-type (including also lenticular galaxies), while those with T-Type $> 0$ are late-type. Fig.~\ref{fig:morphology} shows the distribution of the morphological types of the total sample. We are also interested in late-type galaxies with prominent bulges, which, according to \citet{2018MNRAS.476.3661D} discussion, correspond to galaxies whose probability of hosting a predominant bulge ($P_\mathrm{bulge}$) is greater than 0.8. We additionally identify barred spiral galaxies as those objects whose bar probabilities ($P_\mathrm{bar}$), as obtained from GZ2, exceed 0.8. 

 In this work, we use the star formation histories (SFHs) derived in \citet{2023Natur.619..269D}, obtained through a full-spectral fitting approach using the STECKMAP code \citep{2006MNRAS.365...74O,2006MNRAS.365...46O} on the galaxy central spectra from the SDSS DR16 \citep{2020ApJS..249....3A}. STECKMAP is a non-parametric tool designed to recover both the kinematics and stellar population parameters from integrated light spectra by fitting a combination of Single Stellar Populations (SSPs) to the data. To accurately recover the stellar populations, emission lines were removed from the spectra, and a fixed line-of-sight velocity distribution was assumed. These preprocessing steps were carried out using pPXF \citep{2017MNRAS.466..798C}.

Once the SSPs were fitted to the spectra, we derived key parameters such as the stellar mass fraction, metallicity, and the age of the currently living stars in each galaxy. A correction factor to obtain the stellar mass fraction at a given age, dependent on age and metallicity, was applied following the prescriptions of \citet{2015MNRAS.449.1177V}, using BaSTI isochrones \citep{2004ApJ...612..168P} and the Kroupa \citep{2001MNRAS.322..231K} IMF.

From the corrected stellar mass fractions and stellar population ages, we reconstructed the SFH as the cumulative stellar mass fraction formed over different look-back times. The uncertainties were estimated using the standard deviation of five Monte Carlo realizations of STECKMAP. To determine the timescales for 50\% ($\rm T_{50}$), 70\% ($\rm T_{70}$), and 90\% ($\rm T_{90}$) of stellar mass formation\footnote{The times provided refer to look-back times. For example, a $\rm T_{50} = 8~Gyr$ indicates that the galaxy formed half of its mass 8 billion years ago.}, we interpolated the cumulative SFH.

\section{The stellar mass-size relation}
\label{results}
Figure~\ref{fig:scatterplot_R50_all} shows the stellar mass-size distribution of the final sample, divided in early-type (top panel), and late-type galaxies (bottom panel).  In this section, we will differentiate the results for early- and late-type galaxies for clarity.

We have characterised the stellar mass-size distributions for the galaxies in different environments with a function fitting to the average values. Previous works \citep[e.g][]{2003MNRAS.343..978S,2019RAA....19....6Z} have parametrised the mass-size distribution with an equation of the following type:
\begin{equation}
 \rm R_{50} = \gamma \left( \frac{M_{\star}}{M_{o}} \right)^{\alpha} \left( 1+ \frac{M_{\star}}{M_{o}} \right)^{(\beta-\alpha)},
\label{equation:ecuacion} 
\end{equation}
where the $\rm M_{\star}$ is the stellar mass, $\alpha$ and $\beta$ characterise the slope below and above a certain critical mass, $\rm M_{o}$, and $\gamma$ is another fitting parameter related to the scaling. For comparison purposes we are going to adopt this function fitting to our data.

\subsection{Early-type galaxies}
We conducted the Wilcoxon-Mann-Whitney (WMW) and Kolmogorov-Smirnov (KS) tests to compare the distributions of Petrosian R50 radii for early-type galaxies across different large-scale environments. The statistical analysis reveals extremely low p-values for the comparison of void galaxy sizes with filament and cluster galaxies, suggesting that void galaxies come from a different distribution than those in filaments and clusters. The p-value for the comparison between filament and cluster galaxies is close to the critical threshold of 0.05, indicating that these two populations are drawn from similar distributions. When performing these tests for two distinct stellar mass ranges, ($\rm log(M_\star / \mathrm{M_\odot}) > 10.5$ and $9.0 < \rm log(M_\star / \mathrm{M_\odot}) < 10.5$, the results do not change.

The top panel of Fig.~\ref{fig:scatterplot_R50_all} clearly demonstrates that early-type void galaxies (blue symbols) tend to be smaller than their filament and cluster counterparts of the same mass across all mass ranges (green and red symbols, respectively).
We fit equation~\ref{equation:ecuacion} to the median values of the mass-size relation for void galaxies at each mass bin. Allowing all parameters to vary freely did not result in a satisfactory fit. To achieve a better fit for the void galaxy data, we fixed the critical mass, $\rm M_{o}$, to two values corresponding to the mass regimes outlined in \citet{2019RAA....19....6Z}: the 'high' $\rm M_{o}$ and 'low' $\rm M_{o}$, as listed in Table~\ref{tab:fits}, in this way we ensure the parameter coverage presented in previous works. The resulting fitted values are consistent with those reported by other studies \citep[e.g.][]{2019ApJ...872L..13M}. The results of these fits for the void galaxies are shown in Table~\ref{tab:fits} and in the top panel of Fig.~\ref{fig:scatterplot_R50_all}. Table~\ref{tab:fits} also provides the fitted parameters for elliptical and spiral galaxies from \citet{2019RAA....19....6Z}. 

The slope,  $\alpha$, derived for the 'high' $\rm M_{o}$ is consistent with the values reported by \citet{2019RAA....19....6Z} for spiral galaxies (see Table~\ref{tab:fits}). However, the $\gamma$ parameter is significantly lower than that for spiral galaxies, which is expected since early-type galaxies are inherently smaller, and $\gamma$ serves as a scaling factor. The parameters fitted for the 'low' $\rm M_{o}$ closely match those found for elliptical galaxies by the same group, though with lower, $\beta$ and $\gamma$ values (see Table~\ref{tab:fits}). The results are also consistent with other previous works \citep[e.g.][]{2012MNRAS.422L..62C,2012ApJ...746..162N}. Overall, these results suggest that void galaxies have smaller R50 radii than those in denser environments by about 10-20\% across all masses, with average values provided in Table~\ref{tab:averages}, and that higher-mass early-type void galaxies exhibit a shallower slope compared to early-type galaxies in denser environments.
 
\begin{table}
\caption{The table presents the best-fit parameters for the observed mass-size distribution of galaxies in voids, as derived in this work, alongside the best-fit parameters for the mass-size distribution of elliptical and spiral galaxies from \citet{2019RAA....19....6Z}.}
\centering
\begin{tabular}{lcccc}
\hline
Type - This Work &$\rm M_{o}$ & $\alpha$  &$\beta$ &  $\gamma$ \\
\hline
High $\rm M_{o}$ - Early-type&$4.0\times10^{11}$ &  0.2  & 0.7 &  4.5\\
Low $\rm M_{o}$ - Early-type &$1.2\times10^{10}$&  0.1  &  0.5&  1.5\\
Late-type &$1.2\times10^{12}$&  0.2  &  0.1 &  8.0 \\
\hline
Type - Zhang \& Yang (2019)&$\rm M_{o}$ & $\alpha$  &$\beta$ &  $\gamma$ \\
\hline
Elliptical&$1.3\times10^{10}$ &  0.1  & 0.6 &  1.7\\
Spiral&$1.9\times10^{12}$&  0.2  &  5.4 &  9.0 \\
\hline

\end{tabular}
\label{tab:fits}
\end{table}


\subsection{Late-type galaxies}

We carried out WMW and KS tests to compare the distributions of Petrosian R50 radii for late-type galaxies across different large-scale environments. Statistically significant differences in Petrosian R50 distributions were found for all three pairs of environments (Void-Filament, Void-Cluster, and Filament-Cluster. The most pronounced differences were observed between Filament vs. Cluster and Void vs. Filament, as indicated by their extremely small p-values. While the difference between Void vs. Cluster was also significant, its p-value suggests the evidence is less robust compared to the other pairs.

When repeating these tests for two different stellar mass ranges, ($\rm log(M_\star / \mathrm{M_\odot}) > 10.5$ and $9.0 < \rm log(M_\star / \mathrm{M_\odot}) < 10.5$, the results vary.

In the high-mass range, the p-values for Void-Cluster comparisons, obtained from both the KS and WMW tests, are 0.12. This suggests that these galaxies are likely drawn from the same distribution. However, the p-values for Void-Filament and Filament-Cluster comparisons are extremely small, indicating that these groups are drawn from different distributions.

In the low-mass range, the p-value for the Void-Filament comparison is close to the critical threshold of 0.05, while the p-values for the other two comparisons are extremely small. These findings suggest that cluster galaxies in this mass range are drawn from a different distribution, with cluster galaxies tending to be smaller. As shown in the average values presented in Table~\ref{tab:averages}, late-type cluster galaxies are approximately 10\% smaller than their counterparts in filaments and voids.

\begin{figure}
\centering 
\includegraphics[width = 0.49\textwidth]{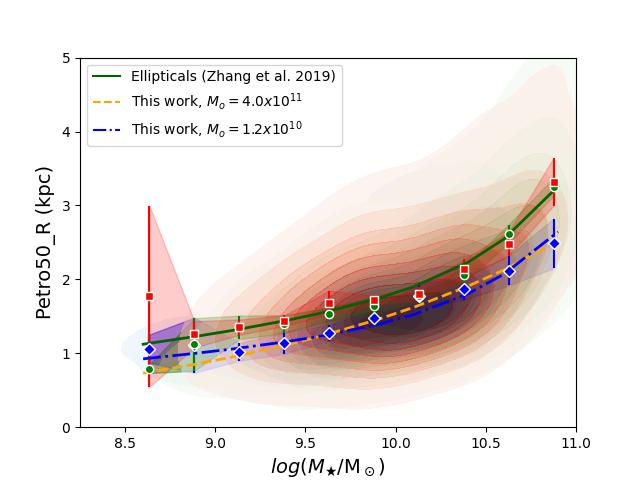}  \includegraphics[width = 0.49\textwidth]{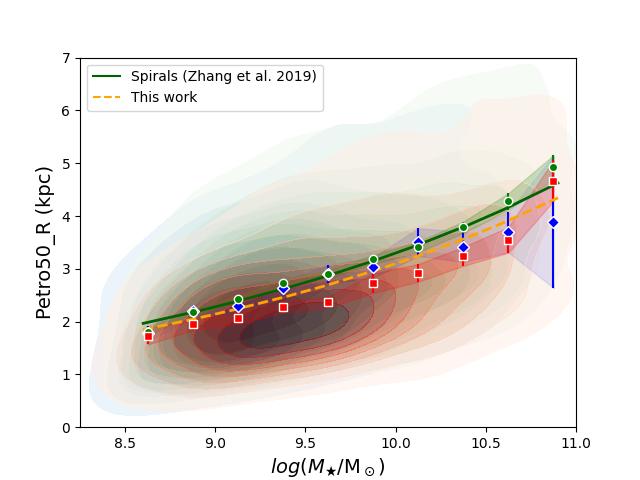}
\caption{The plots illustrate the density and mean values of the Petrosian R50 for early-type galaxies (top panel) and late-type galaxies (bottom panel) across various large-scale environments. Void galaxies are represented in blue, filament galaxies in green, and cluster galaxies in red. For comparison, the green solid lines show the fits from \citet{2019RAA....19....6Z}. In the top panel, the best-fit parameters to the void galaxies are displayed, with blue and orange solid lines in the top panel and an orange solid line in the bottom panel, corresponding to the fitted parameters detailed in Table~\ref{tab:fits}, as discussed in the text.}
\label{fig:scatterplot_R50_all} 
\end{figure}

\begin{figure}
\centering 
\includegraphics[width = 0.49\textwidth]{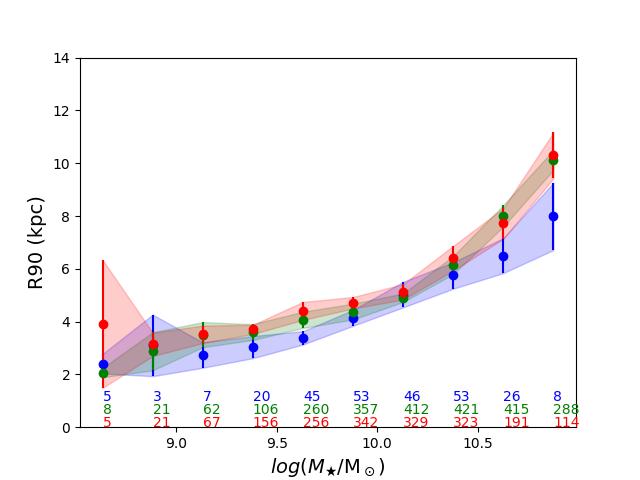}  
\includegraphics[width = 0.49\textwidth]{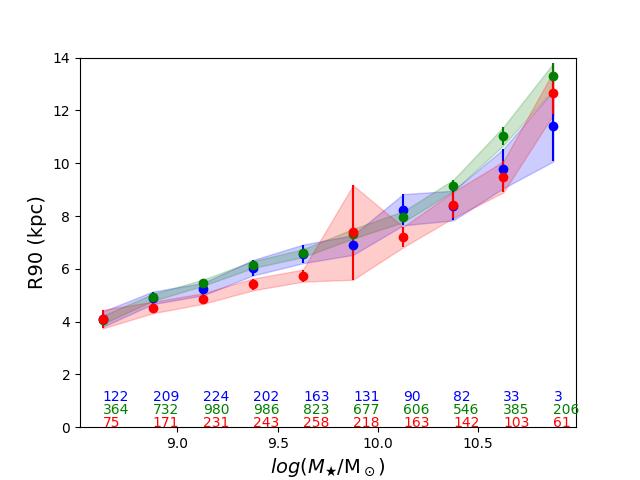}
\caption{The plots display the mean values of the Petrosian R90 for early-type galaxies (top panel) and late-type galaxies (bottom panel) across different large-scale environments and mass bins. Void galaxies are shown in blue, filament galaxies in green, and cluster galaxies in red. The number of galaxies in each mass bin is indicated at the bottom of each panel.}
\label{fig:scatterplot_R90_all} 
\end{figure}

The lower panel of Fig.~\ref{fig:scatterplot_R50_all} shows that for masses above $\rm \approx 10^{10.5} {M_\odot}$, void galaxies have sizes comparable to cluster galaxies, making them smaller than galaxies in filaments, this is corroborated by the KS and WMW p-values for this comparison. 
These findings are consistent when using the Petrosian R90 sizes for the mass-size relation, as shown in Fig.~\ref{fig:scatterplot_R90_all}, and the statistical tests yield similar conclusions.
Figure~\ref{fig:scatterplot_R50_all} also includes the fit to our data, using the functional form described earlier (equ.~\ref{equation:ecuacion}), with all parameters free, alongside the fit from \citet{2019RAA....19....6Z} for their sample, as summarised in Table~\ref{tab:fits}.

\begin{figure}
\centering 
\includegraphics[width = 0.49\textwidth]{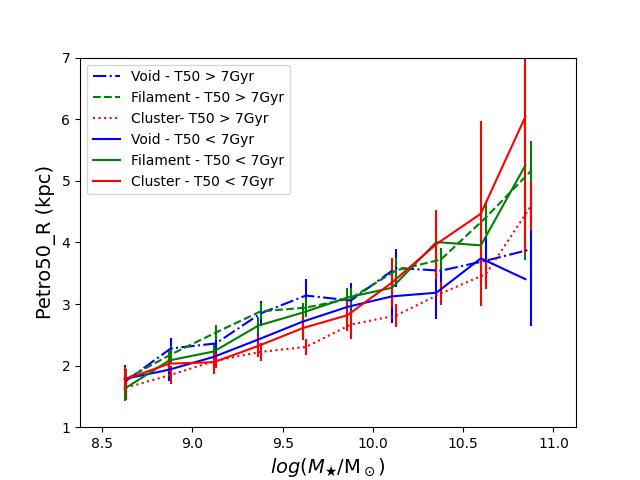} 
\caption{The plot presents the mean stellar mass-size relation, based on the Petrosian R50, for late-type galaxies, distinguishing between those with $T_{50}$ greater than 7 Gyr and those with $T_{50}$ less than 7 Gyr. Void galaxies are represented in blue, filament galaxies in green, and cluster galaxies in red.}
\label{fig:scatterage_R50_lateyoungold} 
\end{figure}

To investigate whether star formation history influences the mass-size relation, we divided late-type galaxies into two groups based on their $\rm T_{50}$ values: above and below 7 Gyr. This threshold was chosen based on simulations suggesting it represents the transition between distinct accretion regimes \citep[e.g.,][]{2006MNRAS.368....2D,2013ApJ...770...57B}. Figure~\ref{fig:scatterage_R50_lateyoungold} illustrates the mass-size relation for these galaxies under this selection criterion.

For late-type galaxies with $\rm T_{50} > 7$ Gyr, cluster galaxies are systematically smaller within the stellar mass range $\rm log(M_\star / \mathrm{M_\odot}) = 9 - 10.5$. This is corroborated by the results of the WMW and KS tests, where the Void-Filament comparison yields a p-value of 0.5, indicating similar distributions. In contrast, the p-values for Void-Cluster and Filament-Cluster comparisons are extremely small, signifying distinct distributions. For galaxies with $\rm T_{50} < 7$ Gyr, in the same mass-range as before, the WMW and KS tests show a large p-value (0.7) for Void-Cluster comparisons, indicating similar distributions, with void and cluster galaxies showing smaller sizes than galaxies residing in filaments. However, the Void-Filament and Filament-Cluster comparisons yield p-values around 0.002, suggesting differences in distributions, although these are less significant than those observed for galaxies with $\rm T_{50} > 7$ Gyr. It is interesting to notice that the analysis of the  $\rm T_{50}$ values for the late-type galaxies yields to different size distributions for the galaxies located in different large-scale environments.

\begin{table*}
\caption{The table shows the average R50 differences for galaxies in different large-scale structures, categorised by early- and late-type galaxies, as well as galaxies with $T_{50}$ above and below 7 Gyr. The columns include the average values and the range of values for each category. }
\centering
\begin{tabular}{llllccc}
\hline
Morph. type & $\rm T_{50}$  &$\rm R50_{fil}-R50_{void}$ &  $\rm R50_{clust}-R50_{void}$ & \# void objects & \# filament objects & \# cluster objects\\
&&[kpc]&[kpc]&\\
\hline
Early-type &   none   & 0.25 [0.04; 0.5] &  0.26 [0.01; 0.4]& 266 &2351& 1804\\
Late-type  &    none     &   0.14 [$-$0.04; 0.7]   &  $-$0.3 [$-$0.6; $-$0.1] & 1259 &6306 & 1665 \\
Early-type  &  > 7~Gyr     &  0.25 [$-$0.02; 0.5]&  0.27 [$-$0.4; $-$0.02] &206&885&1689 \\
Early-type  &  < 7~Gyr      &    -    & - &56&131&80  \\
Late-type  &  > 7~Gyr      & 0.1 [$-$0.1; 0.8]   &   $-$0.45 [$-$0.8; $-$0.01]&706&1609& 1183\\
Late-type  &  < 7~Gyr     & 0.2 [$-$0.15; 0.2]  &  0.1 [$-$0.01; 0.8] &509&1044&432 \\
\hline
\end{tabular}
\label{tab:averages}
\end{table*}
\subsection{Presence of bar and bulges}
\begin{figure}
\centering 
\includegraphics[width = 0.49\textwidth]{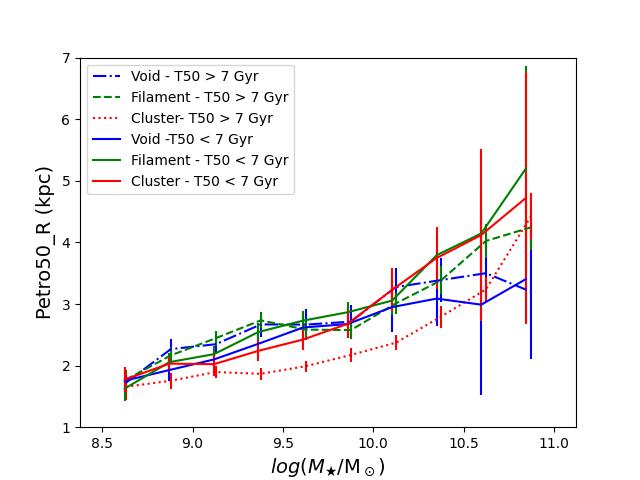} 
\caption{The plot illustrates the mean stellar mass-size relation, based on the Petrosian R50, for galaxies without prominent bulges. Void galaxies are represented in blue, filament galaxies in green, and cluster galaxies in red.}
\label{fig:scatterage_nobulges} 
\end{figure}

Internal structures such as bars and bulges can influence the light distribution of disc galaxies in different ways, making it valuable to characterise the sample of galaxies with prominent bars and bulges. Using the morphological catalogue from \citet{2018MNRAS.476.3661D}, we select galaxies based on their $P_\mathrm{bulge}$ and $P_\mathrm{bar}$ parameters, where $P_\mathrm{bulge}$ represents the probability of hosting a prominent bulge and $P_\mathrm{bar}$ indicates the likelihood of containing a strong bar. Following \citet{2018MNRAS.476.3661D}, we use $P_\mathrm{bar} > 0.8$ as a threshold to identify barred galaxies and $P_\mathrm{bulge} > 0.8$ to select galaxies with significant bulges.

As shown in Fig.\ref{fig:scatterage_nobulges}, the mass-size distribution of galaxies without bulges follows a similar trend to that observed for late-type galaxies. Specifically, for bulgeless galaxies with $\rm T_{50} > 7 Gyr$, those in clusters are smaller than their counterparts in lower-density environments. This observation is supported by the WMW and KS analysis, where the p-values in the Void-Filament are around the limit value of 0.05 for both tests, while the Void-Cluster and the Filament-Cluster comparison p-values are extremely small. In contrast, bulgeless galaxies with $\rm T_{50} < 7~Gyr$ and masses below $\rm \approx 10^{10.5} {M_\odot}$ display no significant variation in their mass-size distribution based on their location within the large-scale structure as shown by the large p-value results (p-value=0.1) of the KS and WMW tests on the comparison of all samples.

The fraction of void galaxies with $P_{bulge} > 0.8$ is less than 10\% (123 galaxies) and spread mostly between $\rm log(M_{\star} / \mathrm{M_\odot}) = 10 - 10.5$, so the conclusions regarding the mass-size distribution for galaxies hosting bulges and environment are not statistically robust. However, with the current data, there are indications that galaxies with bulges have similar R50 and R90 independently of their current location in the large-scale environment. The populations of bulged galaxies is concentrated (for all environments) on galaxies with $\rm T_{50} > 7 Gyr$. Only a very small fraction of galaxies with bulges have $\rm T_{50} < 7 Gyr$ (15 galaxies in voids, 38 galaxies in filaments and 18 galaxies in clusters).  

The distributions of R50 and R90 for barred galaxies do not show significant differences across various large-scale structure environments. However, galaxies in clusters may appear slightly smaller, consistent with the trend observed for late-type galaxies, as indicated by the KS and WMW test results applied to the distributions. It is important to note that the sample size is limited, with only 64 barred galaxies identified in voids, making it necessary to confirm these findings with a larger dataset.
 
\subsection{ST- vs LT-SFHs}
\label{subsec:LT_ST}

\begin{figure}
\centering 
\includegraphics[width = 0.49\textwidth]{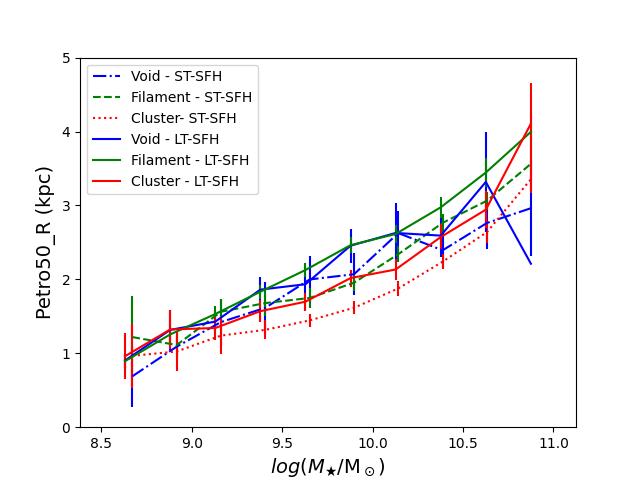} 
\caption{The plot shows the mean stellar mass-size relation, using the Petrosian R50, for galaxies with short-timescale (ST-SFH) and long-timescale star formation histories (LT-SFH). Void galaxies are represented in blue, filament galaxies in green, and cluster galaxies in red.}
\label{fig:scatterage_LT_ST} 
\end{figure}

To investigate whether the galaxy size of the full sample depends on star formation history (SFH), we examine the bimodal distribution of SFHs identified in \citet{2023Natur.619..269D}, which led to a classification of galaxies into short-timescale SFHs (ST-SFH) and long-timescale SFHs (LT-SFH)\footnote{The short-timescale star formation history (ST-SFH) is defined in \citet{2023Natur.619..269D} as the SFH where around 30\% of the total stellar mass formed around 12~Gyr ago. In contrast, the long-timescale star formation history (LT-SFH) exhibits a more gradual and consistent rate of star formation spread out over time.}. 

Figure~\ref{fig:scatterage_LT_ST} illustrates the size distribution of galaxies categorised by their star formation histories (SFH) as explained above. We began our statistical analysis by applying the WMW and KS tests to examine the size distributions of LT-SFH and ST-SFH galaxies across the three large-scale environments. In voids, LT-SFH and ST-SFH galaxies exhibit similar size distributions, as indicated by the high p-value (0.9). In contrast, the size distributions of LT-SFH and ST-SFH galaxies differ significantly in other environments, evidenced by the extremely low p-values. These results suggest that the impact of SFH on galaxy sizes is negligible in cosmic voids, whereas it appears to play a more significant role in other large-scale environments.

The WMW and KS statistical tests performed on the size distributions of LT-SFH and ST-SFH galaxies, separately, within the mass range $\rm log(M_\star / \mathrm{M_\odot}) = 9 - 10.5$, reveal no significant differences between filament and void galaxies, with p-values around 0.5. However, comparisons involving cluster galaxies yield extremely low p-values, indicating that cluster galaxies in this mass range are smaller than their counterparts in filaments and voids for both LT- and ST-SFH galaxies. In \citet{2023Natur.619..269D} the authors discussed the relationship between this classification and the colour and morphology of galaxies, noting that while there is some correlation, it is far from a one-to-one correspondence. Figure \ref{fig:sfhdist} illustrates the distribution of the two SFH types on the colour-magnitude diagram across different large-scale environments. Both SFH types occupy similar parameter spaces in various environments, which needs to be taken into account to explain the correlation found here.

\subsection{Concentration parameter}

\begin{figure}
\centering 
\includegraphics[width = 0.49\textwidth]{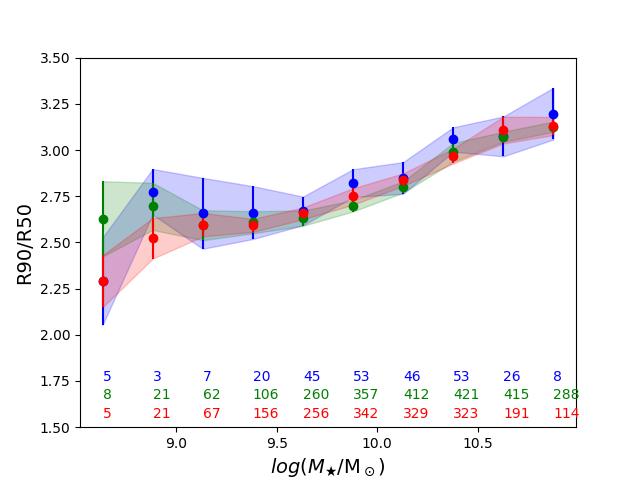}  
\includegraphics[width = 0.49\textwidth]{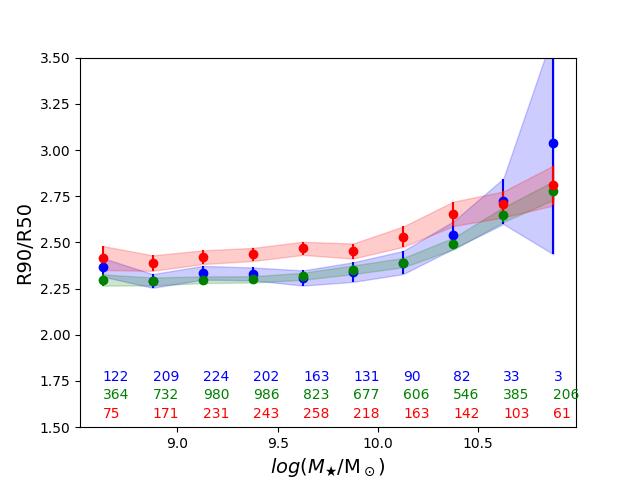}
\caption{The plots show the mean values of the stellar mass-concentration parameter for early-type galaxies (top panel) and late-type galaxies (bottom panel) across different large-scale environments. Void galaxies are represented in blue, filament galaxies in green, and cluster galaxies in red. The number of galaxies in each mass bin is indicated at the bottom of each panel.}
\label{fig:appendixcFig} 
\end{figure}

\begin{figure*}
\centering 
\includegraphics[width = 0.49\textwidth]{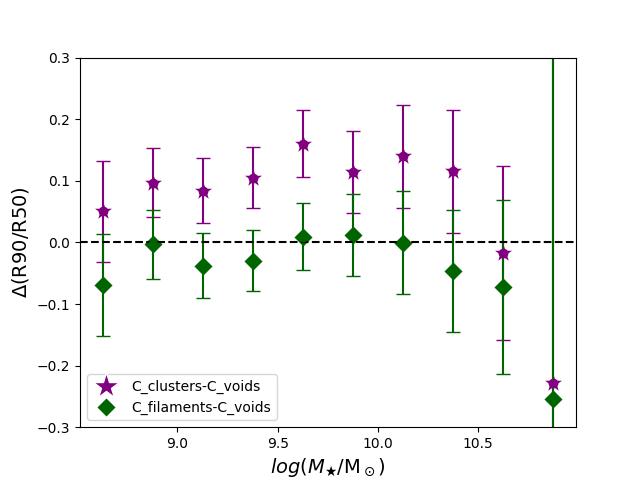}
\includegraphics[width = 0.49\textwidth]{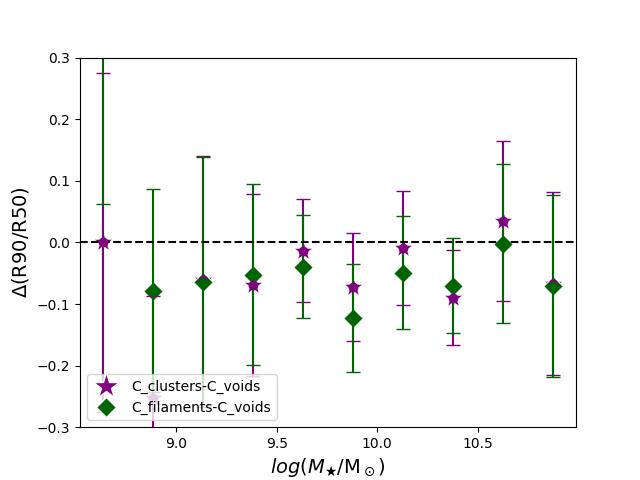} 
\caption{The plots display the difference in the mean concentration ratio (C), defined as R90/R50, at different mass bins between filament and cluster galaxies in comparison to void galaxies. The left panel illustrates this difference for late-type galaxies, while the right panel presents the comparison for early-type galaxies.}
\label{fig:ratio_late} 
\end{figure*}

The concentration parameter is defined as the ratio between the R90 and R50 radii, and is related to how concentrated the light distribution in a galaxy is: a larger concentration parameter indicates a less concentrated light distribution.
The bottom panel of Fig.~\ref{fig:appendixcFig} shows the distribution of the concentration parameter for late-type galaxies, the ratio remains relatively constant up to a stellar mass of $\rm log(M_\star / \mathrm{M_\odot}) = 10$, with an average value of 2.3. Beyond this mass, the ratio increases, reaching 3.0 for the most massive galaxies (see Fig.~\ref{fig:ratio_late}). For cluster galaxies, the average concentration parameter is 2.45 up to $\rm log(M_\star / \mathrm{M_\odot}) = 10$, which is approximately 10\% higher than that of galaxies in voids and filaments, a trend clearly visible in Fig.~\ref{fig:appendixcFig}. Statistical tests (KS and WMW) show that the concentration parameter distribution for void and filament galaxies is similar (with KS and WMW p-values around 0.4). In contrast, comparisons with cluster galaxies yield very low p-values, indicating that the concentration parameter distributions for cluster galaxies are significantly different from those in void and filament environments. 

The top panel of Fig.~\ref{fig:appendixcFig} shows the distribution of the concentration parameter for early-type galaxies. The concentration increases linearly with stellar mass, with similar slopes observed for galaxies across different large-scale environments. Statistical analysis of the concentration parameter distribution using WMW and KS tests reveals p-values close to or well above the critical value of 0.05, leading to the conclusion that there is no significant difference in the concentration parameter of early-type galaxies across different large-scale environments.

Another way to visualise the differences in the concentration parameter across different large-scale environments is shown in Fig.~\ref{fig:ratio_late}, where the difference in average concentration values is plotted for various mass bins across the different environments. Differences above zero indicate galaxies that are more compact than void galaxies. Analysis of the KS and WMW tests comparing the differences for early-type galaxies yields p-values around 1 for both the Cluster-Void and Filament-Void comparisons, suggesting that the distributions are similar. However, the comparison of the concentration parameter for late-type galaxies results in p-values of 0.002 for both the Cluster-Void and Filament-Void comparisons, indicating a significant difference in their distributions.

We can conclude that Late-type cluster galaxies are more concentrated than late-type galaxies living in lower density environments. Void and filament galaxies exhibiting similar concentration parameters.


To check that we obtain values comparable to previous works, we have qualitatively compared our results with previous values found in the literature \citep[e.g., ][]{2001AJ....122.1861S,2001AJ....122.1238S,2021A&A...648A.122V}. Our findings and average values agree with their results, where they showed that the morphological types are closely linked to the concentration parameter, or in their case, to the inverse of the concentration parameter. 
For completeness, we have analysed the potential relationship between stellar mass and the bulge fraction (B/T) as derived by \citet{2011ApJS..196...11S} for all galaxies in our sample. Following the methodology outlined by \citet{2011ApJS..196...11S}, we selected only those galaxies that can be adequately described by a bulge-and-disk model. Specifically, we required that the probability 
$P_{\it ps}$, which indicates that a bulge+disk model is not necessary, be less than 0.32. This selection resulted in a significantly reduced sample consisting of 452 void galaxies, 3139 filament galaxies, and 1669 cluster galaxies.

Dividing this sample further into early- and late-type galaxies reduces the sample sizes even more. Despite the smaller sample sizes, the observed trends of B/T with stellar mass align closely with those shown in Fig.~\ref{fig:appendixcFig}. Specifically, early-type galaxies exhibit an increasing B/T ratio with stellar mass. Void galaxies tend to have larger B/T values at a given stellar mass compared to their counterparts in denser large-scale environments. Meanwhile, late-type galaxies show a more linear trend of B/T with stellar mass, with cluster galaxies exhibiting higher B/T values at a given stellar mass compared to other environments.
However, a more detailed analysis of the structural properties and larger sample sizes are required to draw robust conclusions about the influence of the large-scale environment on the bulge fraction of galaxies.

\section{Discussion} \label{sec:discussion}
\label{discussion}
Before contextualising the results presented in the previous sections and their implications for galaxy evolution, it is important to first address the potential caveats associated with the analysis.
The main problem with the analysis presented here might come from problems in the estimation of the Petrosian R90 that may produce misleading concentration parameters. One well known, but not well characterised, effect is the fact that neither the R50 or the R90 are corrected for the seeing; this could cause an overestimation of the R90/R50 for those galaxies of size comparable to the point spread function. The average SDSS images seeing in the $\rm r$-band is 1.3 arcsec. Less than 2\% of the total sample have R50 comparable with the average seeing, so we do not expect this effect to affect in any way the results shown here. 
For low-surface brightness extended objects, shallow data would imply an underestimation of the R90 radius that would artificially increase the concentration. This effect will be more important in the case of low mass galaxies. Fig.~\ref{fig:appendixcFig} shows an increase of the R90/R50 value for early-type galaxies with masses below $\rm log(M_\star / \mathrm{M_\odot}) = 9.5$. Apart from the poor statistics in this regime, the effect discussed above might be also playing a role.  These low-mass early type galaxies tend to be of lower luminosity compared to star-forming galaxies of the same mass \citep[see for instance, ][]{2017MNRAS.468.4039R,2021A&A...649L..14R}, as their younger (bluer) populations are more detectable with the SDSS images. This effect is expected to be less strong for the late-type galaxies. 

As discussed in Sect.~\ref{sec:methodology}, a more accurate measurement of galaxy size would follow the approach outlined by \citet{2020MNRAS.493...87T}, which defines galaxy size based on the radial location of the gas surface mass density threshold for star formation. However, to enable comparisons with other studies and due to the SDSS images lacking the necessary surface brightness depth for such measurements, we have opted to use R50 and R90 as proxies for galaxy size in this analysis.

Dust can potentially influence the estimation of structural galaxy parameters, with varying effects depending on the morphological type of the galaxy \citep[e.g.][]{2020MNRAS.493.3580P}. However, recent studies on the molecular gas content of galaxies in voids \citep{2024A&A...692A.125R} have concluded that the overall gas content and specific star formation rates of void galaxies do not differ significantly from those in denser environments. As a result, the impact of dust is expected to be similar across galaxies in different large-scale environments, making it a less concerning factor in the current comparison.

Regarding the mass-size relation, from the results shown in the previous sections, there is clearly an influence of the current location of a galaxy in the cosmic web and on the extent, and distribution, of its stellar component. To our knowledge, this is the first systematic study on the sizes of galaxies in cosmic voids. Previous works focused on under-dense environments have looked at the topic from a local-scale perspective, basing their results on isolated galaxies. This hampers a direct comparison of our results with the literature since, based on the isolation criteria of \citet{2015A&A...578A.110A}, only 10\% of the galaxies populating cosmic voids from the \citet{2012MNRAS.421..926P} catalogue (with z<0.08) qualify as isolated. Consequently, a comparison with isolated objects across different large-scale environments would inadvertently combine large- and local-scale environmental effects in an uncontrolled manner.

Our analysis reveal that early-type galaxies up to $\rm log(M_\star / \mathrm{M_\odot}) = 10$ residing in voids are between 10 and 20\% smaller than galaxies residing in denser environments for the same mass range. There is no clear difference for void galaxies being more compact than galaxies in filaments and clusters, although there is small indication that for early-type galaxies, void objects might be slightly more extended than galaxies living in denser environments. 
Numerical simulations have shown that the mass-size growth rate of galaxies depends on their stellar mass and the time at which they become quenched and their size up to that moment \citep[e.g.][]{2018MNRAS.474.3976G}. This two-phase formation of early-type galaxies in which most of the mass is formed by star formation occurring at early times with gas-rich mergers, followed by a further mass assembly from dry (or almost dry) mergers, has been widely discussed in the literature \citep[e.g.][]{2010ApJ...725.2312O,2012MNRAS.425..641L}. Numerical works have shown that in this scenario, the accreted stars are placed at larger galactocentric radii, showing lower concentrations, than those stars that formed within the galaxy \citep{2013MNRAS.428.3121M,2015ApJ...799..184P,2016MNRAS.458.2371R}.
 
The relation we find regarding the smaller size of early-type void galaxies could be explained if most of the stellar mass is built before quenching time. After this, in denser environments dry minor merger would drive  galaxies to grow more in size that in mass \citep{2017MNRAS.470..651R}, at this stage a steep post-quenching size-mass relation is expected \citep{2015ApJ...813...23V,2018MNRAS.474.3976G}. In cosmic voids, the quenching time occurs later in time \citep{2023A&A...680A.111D} and less minor mergers are expected, both effects could be driving the relation. The fact that the slope of the relation is shallower for void galaxies compared to those galaxies living in filaments above $\rm log(M_\star / \mathrm{M_\odot}) = 10$ might be indicating that there is indeed a lower rate of minor mergers in void environment after quenching either caused by a {\it delayed} evolution, as shown from the SFHs, or by an intrinsic lower rate of minor mergers. In any case, both would be originated by the extreme low density present in cosmic voids.
In fact, as we move to higher redshifts \citep[][]{2013MNRAS.435..207L} early-type galaxies in denser environments are larger than those living in underdense regions, effect apparently not seen at low redshift. Void galaxies evolve slower than their counterparts located in denser regions which might be mimicking the effect seeing at higher redshift. 
Recent numerical simulations \citep[e.g.,][]{2024ApJ...962...58C} suggest that galaxies in voids are generally smaller than those residing in denser regions. These simulations are based on low-mass/spatial resolution models, such as the TNG300 simulation \citep[e.g.,][]{2018MNRAS.475..624N}, which limits the ability to conduct a detailed analysis of the mass-size relation. Consequently, a direct comparison with our results is not feasible. However, it is worth noting that future high-resolution numerical simulations will provide valuable insights into the mass-size relation for galaxies in voids.

The fact that cluster late-type quenched galaxies seem to be smaller and more concentrated (i.e. larger concentration parameter) than filament and void galaxies might be indicative that their current mass is due to their growth during a period of star formation phase early in the Universe, and their subsequent growth in size was quenched by the environment and no further mass growth happened as, possibly, no more gas was available in the cluster environment with the physical condition to form more stars.  These results are compatible with \citet{2010MNRAS.402..282M}, that from a study of 1200 galaxies of field (notice that there is no division about the large scale environment) and cluster galaxies at intermediate redshift, concluded that there is no environment dependence of galaxy size evolution, although they find that for intermediate mass late-type galaxies, those residing in clusters tend to be 15-20\% smaller than field galaxies. They argue that fragile extended disks do not survive the cluster environment \citep[see also][]{2009MNRAS.398.1129G,2009MNRAS.394.1213W}. A more recent work on the impact of the environment on scale relations of main sequence galaxies \citep{2022A&A...665A..54M}, also finds that galaxies living in dense regions are smaller by 14\% and have star formation rates reduced by a factor 1.3-1.5 with respect to field galaxies at z$~\approx$~0.7, notice that no distinction about the large scale environment is made here. There has been recently argued that red spirals are caused by feedback from the AGN quenching rather than from environmental processes \citep{2022A&A...667A..27L}.

We find here that late-type quenched galaxies living in lower density environments as compared to clusters (i.e. filaments, and voids) show a lower concentration at a given mass bin than the cluster late-type quenched galaxies for the same mass range. This suggests an environmental effect rather than feedback from the black hole as the process dominating the mass-size for these galaxies. The evidence for a higher concentration parameter for these late-type cluster galaxies compared to galaxies in lower density environments might be due to stellar redistribution, mass-loss \citep{2012MNRAS.422.1714N}, also in agreement with \citet{2018MNRAS.474.3976G}, or, as mentioned before, their disks might not have survived the tidal disruptions occurring in clusters. 

 The cluster population of galaxies where their $\rm T_{50}$ is below 7~Gyr (around 25\% of the late-type population in the cluster subsample) does not show this behaviour, their sizes are compatible, within the errors, with those galaxies living in less dense environments. Although these cluster galaxies seem to be slightly more concentrated than their counterparts in lower density environments, the difference is within the 95\%  confidence interval of the distribution. 
 These galaxies might have entered the cluster at later stages and they have still survived the dense environment. In fact, a recent work \citep{2023A&A...672A..84D}, analysing Fornax galaxies with IFU data, concluded that the assembling of cold disks is strongly affected by their infall time into the cluster, with older disks corresponding to earlier infallers. The fact that we only see this size segregation for old quenched disks in the cluster environment points to the cluster as the mechanism that shrinks the disks, and not to certain characteristics of the disk properties before falling into the cluster as argued in other works \citep[e.g,][]{2018ApJ...866...78H,2021A&A...647A.100S}.

It is interesting to notice that above $\rm 10^{10.5} M_\odot$ the size of late-type void galaxies is smaller, and the slope of the relation shallower, than that of galaxies in filaments, we can speculate here that this effect might be related to the lower minor-merger rate occurring in the void environment, this is possibly highlighting the importance of these effects in the growth of galaxies in this mass range in the filament environment. In fact, numerical simulations of galaxies in voids \citep{2022MNRAS.517..712R} show that galaxies located in the inner void have a lower rate of minor mergers above that particular stellar mas than galaxies in other environments. The increase of galaxy size by minor merger accretion has been discussed in previous works \citep[e.g.,][]{2011MNRAS.415.3903T}, and here we are seeing the effect thanks to the extremely low density environment found in cosmic voids. This critical stellar mass (often referred to as the "golden mass") has been previously identified as a key threshold for various galaxy properties, such as star formation rates and black hole masses \citep[e.g.,][]{2012MNRAS.421..621B,2016ARA&A..54..597C,2017A&A...599A..71D,2018A&A...620A.113A}. Simulations have attributed the change in these properties at this mass to the effects of cold accretion \citep{2006MNRAS.368....2D,2013ApJ...770...57B}, which may also play a significant role in galaxies located within cosmic voids.

Independently of the large-scale environment, we find that galaxies with bulges have on average $\rm T_{50} > 7 Gyr$ and the mass-size distribution does not vary with their location on the cosmic web. This points to a bulge formation early in time, where major mergers dominated the galaxy growth in both mass and size \citep[in agreement with numerical simulations, e.g.][]{2010MNRAS.401.1099H,2011MNRAS.414.1439D}.

As mentioned in the introduction, galaxies in cosmic voids contribute approximately 10\% of the total mass in the Universe \citep[e.g.,][]{2018MNRAS.473.1195L}. It is therefore unsurprising that their impact on the mass-size relation has largely gone unnoticed when studying the effects of environment on this scaling relation. If only filament and cluster galaxies were considered, the environmental influence on early-type galaxies would have appeared negligible, as no significant differences are observed between field and cluster galaxies, consistent with previous studies \citep{2010ApJ...715..606N,2013ApJ...778L...2C}. However, it is clear from the work presented here that the large-scale structure places a crucial role in shaping galaxies. The early galaxy growth seems to be more efficient in more massive galaxies \citep{2003MNRAS.343..978S,2004ApJ...600L.107F,2006MNRAS.373L..36T,2014ApJ...788...28V,2013ApJ...777..155O}. In this study, we limit the stellar mass range to $\rm log(M_\star / \mathrm{M_\odot}) > 10^{11}$, as there are no void galaxies above this threshold. Exploring the behavior of filament and cluster galaxies beyond this mass range will be reserved for future work, as some of the differences observed in this study may become even more pronounced at higher stellar masses.

\section{Summary and conclusions}
\label{summary}
 We have explored the mass-size relation for a well defined sample of galaxies residing in cosmic voids, filaments, and clusters, framed in the CAVITY project, with stellar mass ranging between $\rm 10^{8.5} - 10^{11} M_\odot$.  
 The size parametrisation of the data is given by the SDSS Petrosian R50, R90 an their ratio, distinguising between early- and late-type morphologies. We also have information about the percentage of galaxies hosting a bar and a bulge.  From a previous analysis of the SFHs of the central SDSS spectra, we have the times at which the galaxies have assembled 50, 70, and 90\% of their total stellar mass. 

We find the following trends for the mass-size relation of the analysed sample.
 \begin{itemize}
\item Both R50, and R90 are systematically smaller (10-20\%) for early-type void galaxies in all mass bins.
\item The mass-size relation slope is shallower for early-type void galaxies.
\item No significant difference with environment is found in the concentration parameter for early-type galaxies in the mass bins where reliable R90/R50 ratios can be derived.
\item In the mass interval of  $\rm log(M_\star / \mathrm{M_\odot}) = 9 - 10.5$ late-type cluster galaxies tend to be smaller than galaxies in filaments and voids. 
\item Late-type void galaxies with masses above $\rm \approx 10^{10.5} M_\odot$ present a shallower mass-size slope and smaller sizes than galaxies in filaments.
\item For  masses $\rm log(M_\star / \mathrm{M_\odot}) = 9 - 10.5$, late-type cluster galaxies with $\rm T_{50}$ > 7~Gyr are smaller than galaxies in lower density regions within the same mass bin. For late-type galaxies with $\rm T_{50}$ < 7~Gyr, in the same mass range, cluster and void galaxies exhibit similar distributions, with both having statistically smaller sizes compared to galaxies in filaments.
\item Late-type cluster galaxies present higher light concentration than filament and void galaxies for the same mass bin.
\item The impact of SFH on galaxy sizes is negligible in cosmic voids, whereas it appears to play a more significant role in
other large-scale environments.
\end{itemize}

In summary, we present clear evidence of galaxy stellar mass-size segregation in relation to the large-scale environment. Early-type galaxies, which are thought to grow in two distinct phases, primarily increase their mass during the initial gas-rich major-merger stage. In voids, the subsequent size growth through minor mergers is less pronounced, likely due to slower evolutionary processes, reduced minor merger activity, fewer accretion events, or a combination of these factors. The shift in the slope for high-mass void galaxies indicates a lower rate of minor accretion. This pattern is also observed in late-type void galaxies with masses above $\rm \approx 10^{10.5} M_\odot$, where minor mergers contribute to their size growth. In contrast, late-type quenched cluster galaxies are smaller, mainly due to frequent interactions within the cluster environment, with early infallers being particularly affected by these environmental effects.
 In future work, more precise measurements of galaxy sizes will be possible using the deep imaging data from the CAVITY collaboration, which will provide deeper insights into the role of environment in shaping the outer regions of galaxies.


\begin{acknowledgements}
We acknowledge financial support from the research project PRE2021-098736 funded by MCIN/AEI/10.13039/501100011033 and FSE+. We acknowledge financial support by the research projects PID2020-113689GB-I00, PID2023-149578NB-I00, and PID2020-114414GB-I00, financed by MCIN/AEI/10.13039/501100011033, the project A-FQM-510-UGR20 financed from FEDER/Junta de Andalucía-Consejería de Transformación Económica, Industria, Conocimiento y Universidades/Proyecto and by the grants P20\_00334 and FQM108, financed by the Junta de Andalucía (Spain). We also acknowledge financial support from AST22.4.4, funded by Consejería de Universidad, Investigación e Innovación and Gobierno de España and Unión Europea — NextGenerationEU. DE acknowledges support from: 1) a Beatriz Galindo senior fellowship (BG20/00224) from the Spanish Ministry of Science and Innovation,
2) projects PID2020-114414GB-100 and PID2020-113689GB-I00 financed by MCIN/AEI/10.13039/501100011033,
3) project P20-00334  financed by the Junta de Andaluc\'{i}a,
4) project A-FQM-510-UGR20 of the FEDER/Junta de Andaluc\'{i}a-Consejer\'{i}a de Transformaci\'{o}n Econ\'{o}mica, Industria, Conocimiento y Universidades. R.G.B. acknowledges financial support from the Severo Ochoa grant CEX2021-001131-S funded by MCIN/AEI/ 10.13039/501100011033 and to grant PID2022-141755NB-I00. AFM has received support from RYC2021-031099-I and PID2021-123313NA-I00 of MICIN/AEI/10.13039/501100011033/FEDER,UE, NextGenerationEU/PRT. TRL acknowledges support from Juan de la Cierva fellowship (IJC2020-043742-I). JFB acknowledges support from PID2022-140869NB-100. M.A-F. acknowledges support from the Emergia program (EMERGIA20-38888) from Consejer\'ia de Universidad, Investigaci\'on e Innovaci\'on de la Junta de Andaluc\'ia.  J.R. acknowledges financial support from the Spanish Ministry of Science and Innovation through the project PID2022-138896NB-C55. SDP acknowledges financial support from Juan de la Cierva Formaci\'on fellowship (FJC2021-047523-I) financed by MCIN/AEI/10.13039/501100011033 and by the European Union `NextGenerationEU'/PRTR, Ministerio de Econom\'ia y Competitividad under grants PID2019-107408GB-C44, PID2022-136598NB-C32, and is grateful to the Natural Sciences and Engineering Research Council of Canada, the Fonds de Recherche du Qu\'ebec, and the Canada Foundation for Innovation for funding.

This research made use of Astropy, a community-developed core Python (\url{http://www.python.org}) package for astronomy \citep{2022ApJ...935..167A}; ipython \citep{PER-GRA:2007}; matplotlib \citep{Hunter:2007}; SciPy, a collection of open-source software for scientific computing in Python \citep{2020SciPy-NMeth}; and NumPy, a structure for efficient numerical computation \citep{harris2020array}. Funding for the SDSS and SDSS-II has been provided by the Alfred P. Sloan Foundation, the Participating Institutions, the National Science Foundation, the U.S. Department of Energy, the National Aeronautics and Space Administration, the Japanese Monbukagakusho, the Max Planck Society, and the Higher Education Funding Council for England. The SDSS Web Site is \url{http://www.sdss.org/}.
\end{acknowledgements}

\bibliographystyle{aa} 
\bibliography{lettersizes} 

\end{document}